\documentstyle[aps, prb, WIDEspecial]{revtex}
\include{epsf}
\newcommand{\sruo}{Sr$_2$RuO$_4$}
\newcommand{\degrees}[1]{\mbox{$#1{}^{\circ}$}}
\newcommand{\bm}[1]{\mbox{\boldmath $#1$}}

\begin{document}
\draft

\title{Quantum Oscillations and Overcritical Torque Interaction in
  \sruo}

\author{C.~Bergemann and S.~R.~Julian}
\address{Cavendish Laboratory, University of Cambridge,
  Madingley Road, Cambridge~CB3~0HE, United~Kingdom}

\author{A.~P.~Mackenzie and A.~W.~Tyler} 
\address{School of Physics and Astronomy, University of
  Birmingham, Edgbaston, Birmingham~B15~2TT, United~Kingdom}
  
\author{D.~E.~Farrell} 
\address{Physics Department, Case Western
  Reserve University, Cleveland, OH~44106, USA}
        
\author{Y.~Maeno and S.~NishiZaki} 
\address{Department of Physics,
  Graduate School of Science, Kyoto University, Kyoto 606-8502, Japan}

\date{Submitted to Physica C on 27 September 1998}

\maketitle

\begin{abstract}
  \sruo\ is the only known layered perovskite oxide superconductor
  without copper; there is strong evidence for an unconventional
  (``p-wave'') pairing mechanism, and it has recently been shown to
  possess a cylindrical, ``quasi-two-dimensional'' Fermi surface.
  Using \sruo\ as a test case for the detection of quantum
  oscillations with piezoresistive microcantilevers, the piezolever
  torque magnetometry technique was successfully implemented on a
  dilution refrigerator. It was possible to reproduce the
  quantum-oscillatory magnetization data on all three Fermi surface
  sheets, in a crystal of microgram mass. Moreover, an absolute
  estimate of the amplitude of the oscillation is provided.  We also
  investigated the phenomenon of torque interaction which distorts the
  magnetization signal and introduces harmonics and sidebands to the
  dHvA spectrum. As the torque interaction effect grows in strength,
  overcriticality is shown to lead to discrete magnetization jumps and
  to a near-asymptotically damped ``sproing'' effect.
\end{abstract}
\vspace{.5cm}

\narrowtext

\section{Introduction}

Recent discoveries of unconventional electronic behaviour --- such as
high-$T_c$ superconductivity in the cuprates or colossal
magnetoresistance in the manganates --- have focused scientific
interest on the transition metal oxides. To understand the properties
of metallic oxides, it is important to probe the validity of the Fermi
liquid picture and, if possible, to infer the shape of the Fermi
surface and other quasiparticle properties. Measurement of the
quantum-oscillatory part of the magnetization --- the
de~Haas-van~Alphen (dHvA) effect --- is the most direct way to obtain
such information. However, these investigations are difficult as the
materials in question are frequently non-stoichiometric and are often
only available as very small crystallites.

In particular, \sruo\ has aroused great interest as the only known
superconductor with the same (layered perovskite) crystal structure as
the high-$T_c$ cuprates but without copper.\cite{maeno} Moreover, it
represents a possible candidate for p-wave
superconductivity.\cite{julianSr} \sruo\ is one of the very few
complex materials for which there is a good prospect that a deeper
physical understanding will soon be achieved. High-purity
stoichiometric crystals can be prepared, and its normal state Fermi
surface is known to be ``quasi-2D'', consisting of three slightly
warped cylindrical sheets.\cite{sruoQuosc}

Here, we reexamine the dHvA effect in \sruo\ using piezolever
torque magnetometry: a new and ultra-sensitive technique which is
especially suited for small microcrystals of anisotropic materials. It
was possible to obtain {\em absolute\/} estimates for the
quantum-oscillatory magnetization on all three Fermi surface sheets.
The torque interaction effect and its remarkable manifestation in the
``overcritical'' case will also be discussed.

\section{Piezolever Torque Magnetometry}

The fundamental frequency of the quantum-oscillatory torque density
(magnetic torque per unit volume) corresponding to a Fermi surface
cylinder in a quasi-2D metal is\cite{shoenTorque}
\begin{equation}
  \label{LKM}
  \tilde \tau = \frac{e^2 h_{\rm BZ} F B \sin\theta}{2\pi^3 m^\ast}
  \:R\:
  \sin\left( \frac{2\pi F}{B\cos\theta} \right) 
\end{equation}
where $B$ is the magnetic field which is applied at an angle $\theta$
to the low conductivity axis, $h_{\rm BZ}$ is the Brillouin zone
height (i.e.\ the length of the cylinder), $m^\ast$ the on-axis
quasiparticle mass, and $F$ the dHvA frequency corresponding to the
on-axis Fermi cylinder cross-section $A_F$ via \mbox{$F = \hbar A_F/
  2\pi e$}. The number $R$ represents the damping factors related to finite
temperature, sample inhomogeneity, spin-splitting, and the warping of
the cylinder.\cite{shoenTorque}

\begin{figure}
\centerline{\epsfxsize=6.5cm\epsfbox{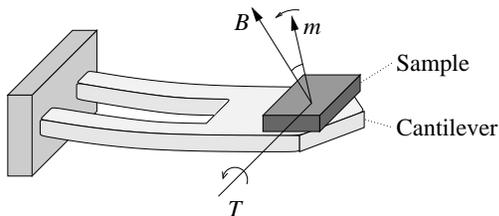}}
\caption{Piezolever torque magnetometry: an applied field 
  \bm{B} induces a magnetic moment \bm{m}, and the magnetic torque
  \mbox{$\bm{T} = \bm{m} \wedge \bm{B}$} in an anisotropic crystal
  flexes the cantilever on which the sample is mounted. This deflection
  is sensed via the piezoresistance of a resistive path implanted on
  the lever.}
\label{simplelever}
\end{figure}

To record this small torque, we have employed piezoresistive
microcantilever torque magnetometry as shown in
Fig.~\ref{simplelever}. This technique was pioneered only
recently;\cite{rossel,ich2} we successfully implemented it at low
temperatures and achieved quantum-oscillatory moment
sensitivities of \mbox{$10^{-12}\,{\rm Am}^2$} at 150\,mK in
a 17\,T field.\cite{ich3} 

A small high-quality \sruo\ crystal ($200 \times 150 \times 20
  \, \mu {\rm m}^3$, residual resistivity about
0.6\,$\mu\Omega\,$cm), prepared in a floating zone image furnace, was
mounted on a piezolever\cite{PSI} of dimensions \mbox{$170 \times 50
  \times 5 \, \mu {\rm m}^3$}. A second (empty) piezolever was used
for background compensation in a Wheatstone bridge circuit, driven by
a 50\,nA amplitude AC current through each lever. Temperatures of
150\,mK were achieved on a dilution cryostat during field sweeps from
18\,T to 15\,T.

\begin{figure}
\centerline{\epsfxsize=8cm\epsfbox{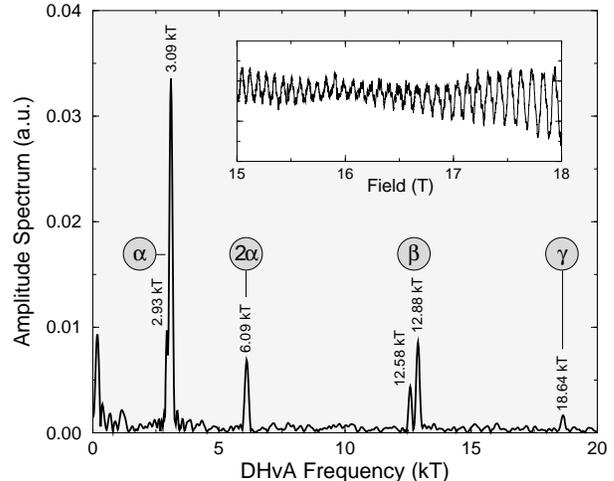}}
\caption{Quantum-oscillatory torque (inset) and resulting dHvA
  spectrum (main panel), as obtained from piezolever torque
  magnetometry. All dHvA frequencies observed in the original
  study\cite{sruoQuosc} (labeled $\alpha$, $\beta$, and $\gamma$) are
  reproduced.}
\label{SrII}
\end{figure}

The output signal and its dHvA spectrum for a sweep close
(\mbox{$\theta \simeq \degrees{2}$}) to the $c$-axis is shown in
Fig.~\ref{SrII}. All dHvA frequencies observed in the original
study\cite{sruoQuosc} (which employed the standard field modulation
technique) were reproduced with piezolever torque magnetometry.

One can extract the {\em absolute\/} quantum-oscillatory
magnetization, and we estimate 460\,A/m, 200\,A/m, and 30\,A/m,
respectively, as the dHvA amplitudes (at \mbox{$T \simeq 0$}, \mbox{$B
  \simeq 17\,$T}, and \mbox{$\theta \simeq \degrees{0}$}) for the
$\alpha$, $\beta$, and $\gamma$ sheets --- to within a factor of two.
These values are in agreement with Eq.~(\ref{LKM}).

\section{Torque Interaction}

\begin{figure}
\centerline{\epsfxsize=18cm\epsfbox{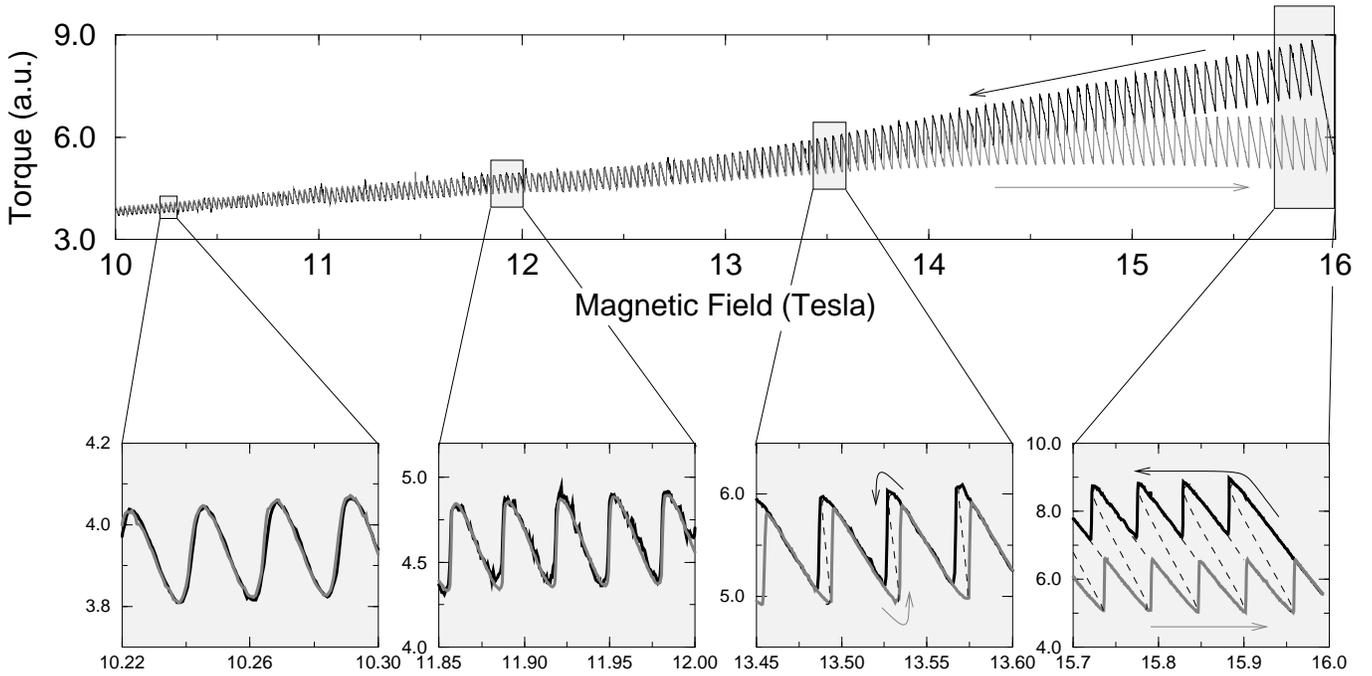}}
\caption{Experimental
  magnetization curve recorded on \sruo, with the magnetic field
  pointing \degrees{48} from the $c$-axis of the crystal. Torque
  interaction shears the sinusoidal profile, and this effect gets
  progressively more severe as the field is swept higher. Above about
  12.5\,T, the effect becomes ``overcritical'', with different values
  of the magnetization during field upsweeps (grey) and downsweeps
  (black), and discrete magnetization jumps can be observed. The
  arrows refer to the sweep direction.}
\label{TIoverlay}
\end{figure}

For off-axis fields, the magnetization signal is affected by the
torque interaction effect which is due to the feedback of the
oscillating magnetic moment on the position of the lever in the
field.\cite{shoenTorque,vanderkooy} As a result, the apparent
magnetization profile gets sheared, and harmonics and sidebands are
introduced to the dHvA spectrum. Fortunately, this effect can be
compensated through numerical treatment of the experimental data. For
large $\theta$ and high fields, however, the effect becomes strong
enough (``overcritical'') to produce discrete jumps in the
magnetization and hence in the position of the lever, see
Fig.~\ref{TIoverlay}.

Although observed previously in arsenic,\cite{vanderkooy} this
remarkably macroscopic manifestation of an inherently quantum
phenomenon was visible in our experiments at unprecedented strength.
The ``sproing'' effect at the magnetization jumps --- when the lever
snaps into its new equilibrium position --- was also recorded with a
high-speed voltmeter: the lever movement turned out to be
near-asymptotically damped by eddy currents. This casts some doubt on
the practicability of proposed ``dynamic mode'' piezolever
measurements,\cite{rossel} at least on metallic samples.
 
\section{Conclusion}

In conclusion, piezolever torque magnetometry measurements have been
able to reproduce the quantum-oscillatory magnetization data on all
three Fermi surface sheets of the layered perovskite oxide
superconductor \sruo. Moreover, they provide an absolute estimate of
the amplitude of the oscillation. This introduces the piezolever
technique as an interesting alternative to conventional methods for
dHvA experiments on anisotropic compounds if these are only available
as microcrystals --- a common situation for complex modern materials.

We also investigated the phenomenon of torque interaction which
distorts the magnetization signal and introduces harmonics and
sidebands to the dHvA spectrum. As torque interaction grows
in strength, it leads to irreversibility effects and discrete
magnetization jumps: a near-asymptotically damped ``sproing'' effect.

We benefited from fruitful discussions with D.~Shoenberg and
M.~J.~Naughton. This work was supported by the U.K.\ EPSRC\@.

\widetext
\narrowtext

\widetext

\end{document}